\documentclass[prl,aps,twocolumn]{revtex4}
\usepackage{amssymb}
\usepackage{amsmath}
\usepackage{epsfig}
\usepackage{t1enc}
\usepackage[polish]{babel}
\usepackage[cp1250]{inputenc}

\begin{document}

\title{Interference of Fock states in a single measurement}

\author{Andrzej Dragan$^1$ and Pawe"{l} Zi"{n}$^2$}

\affiliation{$^1$Institute of Theoretical Physics, Warsaw
University, Ho"{z}a 69, 00-681 Warsaw, Poland\\
$^2$Soltan Institute for Nuclear Studies, Hoża 69, 00-681 Warsaw,
Poland}

\begin{abstract}
We study analytically the structure of an arbitrary order
correlation function for a pair of Fock states and prove without
any approximations that in a single measurement of particle
positions interference effects must occur as experimentally
observed with Bose-Einstein condensates. We also show that the
noise level present in the statistics is slightly lower than for a
respective measurement of phase states.
\end{abstract}

\maketitle

Paul Dirac in his famous textbook on quantum mechanics \cite
{Dirac1967} describes photon interference in the following way.
{\em Suppose we have a beam of light consisting of a large number
of photons split up into two components of equal intensity (...).
If the two components are now made to interfere, we should require
a photon in one component to be able to interfere with one in the
other. Sometimes these two photons would have to annihilate one
another and other times they would have to produce four photons.
This would contradict the conservation of energy (...). Each
photon then interferes only with itself. Interference between two
different photons never occurs.} This view has been criticized
\cite{Pfleegor1967} for applying only to the states of a definite
number of particles. For example two independent sources of
coherent states which have no definite energy can generate an
interference pattern without question.

Dirac's argument, however, may seem to apply at least to the
particle-number Fock states that do have a definite energy value.
Another reason why Fock states seem incapable of interfering is
that they do not have well defined relative phase. And the last
reason - a direct calculation of the first-order correlation
function for two Fock states does not reveal any interference
properties. Unfortunately, these three very attractive arguments
fail.

The first beautiful experimental example of the Fock states
interference has been accomplished with Bose-Einstein condensates
that can be thought of as particle number states. In the
double-condensate experiment by Andrews {\em et al.}
\cite{Andrews1997} the authors proved the existence of
interference fringes in the measurement of positions of condensate
atoms. Results of a similar experiment have been recently reported
in Ref.~\cite{Shin2005}. How is it possible?

The reason is that the first-order correlation function is
attributed to an average (over many realizations) density of
particles, while in the experiments \cite{Andrews1997,Shin2005} we
deal with the results of a single measurement. Quantum mechanics
cannot predict the exact result of a single measurement - it only
predicts average values of certain observables or a probability of
a definite result of a single experiment. But since we are dealing
with a huge number of particles, how about using this many
particle probability distribution to predict a typical particle
density profile in a single experiment \cite{SingleRealization}?
Apparently considering only the first-order correlation function
is not enough to guess the typical density shape and one needs to
take into account also higher-order correlation functions.

This issue has been first addressed in a beautiful work of
Javanainen and Yoo \cite{Javanainen1996} where the authors apply
numerical analysis to the studies of the structure of many
particle probability density distributions. In this and the
subsequent numerical experiment \cite{Cirac1996} exploiting the
laws of quantum mechanics the authors show that two Fock states
can indeed reveal an interference pattern in a single measurement.
Obviously after averaging out over many realizations of the
numerical experiment the interference effects disappear as
expected.

In this Letter we analytically study a nature of the high-order
correlation functions to show directly from their mathematical
structure the existence of interference effects in a single
interference measurement of two Fock states. In our analysis we do
not use any approximations, as the previous authors who attempted
to prove this result analytically \cite{Castin1997,Ashhab2002}
assuming orthogonality of the phase states. This approximation is
questionable when the number of particles measured is of the order
of the total number of particles of the system. We also show an
interesting and unintuitive property of the noise present in the
interference pattern. The noise level turns out to be slightly
lower than in the case of multiple drawn positions with a
probability distribution equal to the interference pattern. We
have shown in a numerical test that the difference is very small
and probably out of reach of any experimental observation. Our
result, however, gives an interesting insight into the structure
of the high-order correlation function.

Consider a set of $d$ identical ideal detectors capable of
counting particles. Let the surface $L$ of the $i$-th detector
placed at the position $x_i$ be described by the characteristic
function $\chi(x-x_i)$ and the annihilation operator $\hat{A}_i$
associated with the mode $L^{-\frac{1}{2}}\chi(x-x_i)$. Let us
assume, that the detectors are spatially separated, i.e. $\int
\chi(x-x_i)\chi(x-x_j) \mbox{d}x=L\delta_{ij}$. We calculate an
average product of the particle counts from all $d$ detectors
\cite{Glauber1963}:
\begin{equation}
I(x_1,\ldots,x_d) = \langle \hat{A}_1^\dagger \hat{A}_1\ldots
\hat{A}_d^\dagger \hat{A}_d\rangle.
\end{equation}
The set of particles measured by the detectors is described by the
field operator $\hat{\Psi}(x)$. We assume, that the occupied modes
are slowly-varying in comparison to the size $L$ of the detectors:
\begin{equation}
\hat{A}_i=\int \frac{1}{\sqrt{L}}\chi(x-x_i)\hat{\Psi}(x)\mbox{d}x
\approx \sqrt{L}\hat{\Psi}(x_i).
\end{equation}
From the above formula follows a connection between average
product of detector counts with the $d$-order correlation
function:
\begin{equation}
\label{averageparticlenumber} I(x_1,\ldots,x_d) = L^d\langle
\hat{\Psi}^\dagger(x_1)\ldots\hat{\Psi}^\dagger(x_d)
\hat{\Psi}(x_1)\ldots\hat{\Psi}(x_d)\rangle.
\end{equation}
If we assume that the detectors' size $L$ is so small that each of
them detects, on average, much less than a single particle then
the average product of the particle counts $I(x_1,\ldots,x_d)$ can
be identified with a probability of detection of exactly one
particle by each detector. Thus the probability density $\varrho$
of localizing the first particle at the position $x_1$, the second
particle at $x_2$, etc., equals:
\begin{widetext}
\begin{eqnarray}
\label{probabilitydensity} \varrho(x_1,\ldots,x_d) &=&
\frac{(N-d)!}{N!}\times
\langle\hat{\Psi}^\dagger(x_1)\ldots\hat{\Psi}^\dagger(x_d)
\hat{\Psi}(x_1)\ldots\hat{\Psi}(x_d)\rangle,
\end{eqnarray}
where $N$ is the total number of particles. Let us notice, that
the probability density (\ref{probabilitydensity}) is defined only
for the states of a definite number of particles. This approach
allows one to interpret the physical meaning of the correlation
function of the order $d$ in two ways. On the one hand it is
proportional to the average product of particle counts of $d$
detectors, on the other hand it is related to the probability
density of localizing exactly one particle by each of $d$ very
small detectors.

As long as the detectors are spatially separated an ordering of
the field operators in the expression
(\ref{averageparticlenumber}) and (\ref{probabilitydensity}) is
defined up to the commutation relation
$[\hat{\Psi}(x),\hat{\Psi}^\dagger(y)]=\nolinebreak\delta(x-y)$.
If one wants to continuously extend the expressions to the case of
$x_i=x_j$ for $i\neq j$ then the field operators must be ordered
normally.

Consider a two-mode quantum state $|n,N-n\rangle$ with the first
mode defined by an arbitrary function $u(x)$ and the second
orthogonal mode by $w(x)$ for $x\in[0,1]$. From the expression
(\ref{probabilitydensity}) we calculate the probability
distribution of localizing all the $N$ particles at positions
$x_1, x_2, \ldots, x_N$:
\begin{eqnarray}
\label{asymetricdensity} \varrho_{|n,N-n\rangle}(x_1,\ldots,x_N)
&=& \binom{N}{n}^{-1}\left| \sum_{\cal P}u(x_{{\cal P}(1)})\ldots
u(x_{{\cal P}(n)})w(x_{{\cal P}(n+1)})\ldots w(x_{{\cal P}(N)})
\right|^2,
\end{eqnarray}
where we sum up over permutations ${\cal P}$ of an $N$-element set
excluding the non-trivial permutations acting separately on the
first $n$ elements of the set and the last $N-n$ elements. We will
consider the case  $N=2n$, when exactly $n$ particles occupy each
mode. Using the formula (\ref{probabilitydensity}) we find that
the probability density of detecting $d$ of $2n$ particles at the
positions $x_1, x_2, \ldots, x_d$ can be expressed with the
probability densities for the asymmetric states
(\ref{asymetricdensity}):
\begin{eqnarray}
\label{symmetricdensity} \varrho_{|n,n\rangle}(x_1,\ldots,x_d) &=&
\binom{2n}{n}^{-1}\sum_{j=1}^d
\Theta(n-j)\Theta(n-d+j)\binom{2n-d}{n-d+j}\binom{d}{j}
\varrho_{|j,d-j\rangle}(x_1,\ldots,x_d),
\end{eqnarray}
where $\Theta(x)$ is the Heaviside's theta function. Binomial
coefficients $\binom{n}{k}=\frac{n!}{k!(n-k)!}$ showing up in the
above expression are for $n\gg 1$ bell-shaped functions of $k$
centered around $k=\frac{n}{2}$ and with a dispersion equal to
$\frac{\sqrt{n}}{2}$. We see that the coefficients
$\binom{2n-d}{n-d+j}$ and $\binom{d}{j}$ attain their maxima for
the same value $j=\frac{d}{2}$, but they are characterized by the
different dispersions of the variable $j$: $\frac{\sqrt{2n-d}}{2}$
and $\frac{\sqrt{d}}{2}$, respectively.

Consider a special case of the probability density
(\ref{symmetricdensity}), with only a small fraction of all
particles being measured, $d\ll n$. In this case the distribution
$\binom{2n-d}{n-d+j}$ is much wider than $\binom{d}{j}$ and we can
replace the former with its maximum value. In this case also the
Heaviside's thetas are equal to the unity and we can skip them. As
a result the expression (\ref{symmetricdensity}) can be written in
the following form:
\begin{eqnarray}
\label{approxdensity} \varrho_{|n,n\rangle}(x_1,\ldots,x_d)
\overset{d\ll n}{\approx} \sum_{j=1}^d 2^{-d}\binom{d}{j}
\varrho_{|j,d-j\rangle}(x_1,\ldots,x_d) =
\int_{-\pi}^\pi\frac{\mbox{d}\phi}{2\pi}
\prod_{i=1}^d\frac{1}{2}\left| u(x_i)+e^{i\phi} w(x_i)\right|^2.
\end{eqnarray}
\end{widetext}
We have managed to express the low-order correlation function for
the highly occupied state $|n,n\rangle$ in an elegant form of an
integral over some positive expression. A similar result has been
shown in Refs.~\cite{Castin1997,Ashhab2002}, however the authors
omit the fact that they actually prove it only for $d\ll n$
because of the limited validity of the approximations used. These
approximations are highly questionable when the number of
particles measured is of the order of the total number of
particles $d \sim 2n$, therefore we are going to prove all the
properties of the high-order correlation functions with no
approximations whatsoever.

It turns out, that the Eq.~(\ref{approxdensity}) tells a lot about
a result of a single measurement of positions of $d$ particles.
According to Born's probabilistic interpretation of quantum
mechanics a result of such measurement - the set of measured
positions $x_1, \ldots, x_d$, corresponds to a result of a single
drawing with the probability density
$\varrho_{|n,n\rangle}(x_1,\ldots,x_d)$. Let us try to predict the
result of such drawing using the following simple lemma based on
Bayes' theorem:

Lemma. {\em If $N$-dimensional probability density $\varrho$ can
be represented in the form $\varrho(x_1,\ldots,x_N)=\int
\mbox{d}\xi\, p(\xi) q(x_1,\ldots,x_N|\xi),$ where $p$ is a
one-dimensional probability density and $q$ is an $N$-dimensional
conditional probability distribution (likelihood) then drawing a
set of random variables $(x_1,\ldots,x_N)$ with the probability
$\varrho$ is equivalent to drawing a random variable $\xi$ with
the density $p$, and then drawing the set of random variables
$(x_1,\ldots,x_N)$ with the density $q$ for the chosen $\xi$.}

Proof. Equivalence of both densities can be shown by proving the
equality of arbitrary moments of the distributions. We will use an
elementary theorem about changing the order of integrals. An
arbitrary moment for the second distribution reads:
\begin{eqnarray}
\int\mbox{d}\xi\,p(\xi) \int \mbox{d}x_1\ldots
\mbox{d}x_N\,x_1^{k_1}\ldots x_N^{k_N}q(x_1,\ldots,x_N|\xi)
\nonumber\\
=\int \mbox{d}x_1\ldots \mbox{d}x_N\, x_1^{k_1}\ldots x_N^{k_N}
\int\mbox{d}\xi\,p(\xi)q(x_1,\ldots,x_N|\xi)\nonumber
\end{eqnarray}
and it is equal to the same moment for the distribution $\varrho$.
As we know, the values of all the moments uniquely determine the
probability distribution. QED.

It follows that the result of a single draw with the probability
density (\ref{approxdensity}) can be achieved by a preliminary
draw of the parameter $\phi$ with a flat distribution, and then by
drawing positions of particles according to the separable density
$\prod_{i=1}^d\frac{1}{2}\left|u(x_i)+e^{i\phi} w(x_i)\right|^2$.
The second draw yields positions centered around maxima of the
one-dimensional function $\frac{1}{2}\left| u(x)+e^{i\phi}
w(x)\right|^2$. If we assume $u(x)=w^*(x)=e^{i\pi x}$, $x\in[0,1]$
then every single measurement reveals the interference fringes
with maxima located randomly each time somewhere else. The meaning
of the interference fringes can be made more precise in the
following way. Suppose that the whole space of possible particle
positions is divided into $D$ small areas of equal length and we
examine how many of the first $d$ particles enter each of these
areas in a single measurement \cite{Saba2005}. Each area is
tightly covered by a set of small detectors constituting, so to
say, a single super-detector. We look at the histograms of the
count statistics of the single measurement - if the sizes of the
considered areas are such that each of them swallows on average a
large number of particles, then each histogram should reproduce
the function $\frac{1}{2}\left| u(x)+e^{i\phi} w(x)\right|^2$ for
some $\phi$.

Let us notice that the last expression in the formula
(\ref{approxdensity}) defines a hidden variables model with the
role of hidden parameter played by $\phi$. Therefore the
uncertainty of the phase $\phi$ attributed to the single
measurement of the small portion $d$ of all particles must be of a
classical nature. Although one can establish a link between a
spin-$\frac{1}{2}$ formalism and parameter $\phi$ our observation
indicates that the considered type of measurement cannot lead to
violation of Bell's inequalities.

We have just shown that for the single measurement of the
relatively small number $d$ of particles belonging to the state
$|n,n\rangle$ one observes the interference fringes. Therefore it
is natural to ask about the result of a similar measurement of all
the $2n$ particles. Below we show that the larger number of
particles is being measured, the fringes of even higher quality
are observed. This agrees with the numerical test \cite{Cirac1996}
and the methodology of the experiment \cite{Saba2005}.

The proof is the following. According to our lemma drawing $d$
positions described by the probability distribution
(\ref{approxdensity}) can be achieved by drawing parameter $\phi$,
and then drawing positions with the conditional probability
distribution $\prod_{i=1}^d\frac{1}{2}\left| u(x_i)+e^{i\phi}
w(x_i)\right|^2$. However, according to the expression
(\ref{approxdensity}) another equivalent method of drawing exists
and is based on drawing first the parameter $j$ described by the
distribution $2^{-d}\binom{d}{j}$ and then drawing positions of
the particles with the probability density
$\varrho_{|j,d-j\rangle}(x_1,\ldots,x_d)$ given by the analytic
formula (\ref{asymetricdensity}). Obviously, both methods of
drawing lead to the same result which, as we know, reveals the
interference patterns of the known shape. The second equality in
(\ref{approxdensity}) indicates that the results of drawing of
positions with the probability distribution
$\varrho_{|j,d-j\rangle}(x_1,\ldots,x_d)$ for the parameter $j$
differing from $\frac{d}{2}$ by not more than a few dispersion
lengths $\frac{\sqrt{d}}{2}$ must reveal the interference effects
every time. Independently of the method of drawing each random
histogram will vary from the ideal shape
$\frac{1}{2}\left|u(x)+e^{i\phi} w(x)\right|^2$ because of
statistical fluctuations. Let us introduce the following measure
of these fluctuations defined for an arbitrary result of the
single drawing. Let the number of counts of the $i$-th
super-detector placed at $x_i$ be denoted with $n_i$. For the
histogram of results $\{n_i\}$ we define the following quantity:
\begin{equation}
\label{noisedefinition} \chi^2 =
\inf_\phi\sum_{i=1}^D\left(n_i-\frac{d}{2D}\left|u(x_i)+e^{i\phi}w(x_i)\right|^2\right)^2,
\end{equation}
where $D$ is the number of the super-detectors. The  above
expression averaged out over many realizations $\overline{\chi^2}$
we will call noise. This noise depends only on the number of
super-detectors and the probability distribution $\varrho$, or
equivalently on the quantum state $\hat{\varrho}$ and the
parameters $d$ and $D$, which we denote as
$\overline{\chi^2}(\hat{\varrho},d,D)$. Therefore the better the
histograms reproduce the shape $\frac{1}{2}\left|u(x)+e^{i\phi}
w(x)\right|^2$ (for some $\phi$) the lower the value of noise
$\overline{\chi^2}$. From the last equality in the formula
(\ref{approxdensity}) we get:
\begin{equation}
\label{noiseequality} \sum_{j=1}^{d} 2^{-d}\binom{d}{j}
\overline{\chi^2}(|j,d-j\rangle,d,D) =
\overline{\chi^2}(|d\rangle_\phi,d,D),
\end{equation}
where $|d\rangle_\phi$ is so called phase state of $d$ particles
occupying the same mode $\frac{1}{\sqrt{2}}\left[u(x)+e^{i\phi}
w(x)\right]$. We have used the fact that the quantity
$\overline{\chi^2}(|d\rangle_\phi,d,D)$ cannot depend on the
selection of $\phi$ and it determines the level of noise for the
histogram of particle positions drawed one by one with the
probability density $\frac{1}{2}\left|u(x)+e^{i\phi}
w(x)\right|^2$. Equation (\ref{noiseequality}) indicates that the
noise level $\overline{\chi^2}(|d\rangle_\phi,d,D)$ is equal to an
average noise level for the states $|j,d-j\rangle$ with the
weights equal to $2^{-d}\binom{d}{j}$.

Unfortunately the level of noise averaged out over all states
$|j,d-j\rangle$ does not uniquely determine the value of noise
$\overline{\chi^2}(|j,d-j\rangle,d,D)$ for the particular $j$.
However we can use a natural assumption that the interference
effects disappear for the asymmetric states $|j,d-j\rangle$. In
the extreme but highly improbable example of the state
$|d,0\rangle$ or $|0,d\rangle$ the interference will be obviously
completely absent. To be more specific, we assume that
$\overline{\chi^2}(|j,d-j\rangle,n,D)$ is a monotonically
increasing function of $\left|j- \frac{d}{2} \right|$. According
to this assumption and the equation (\ref{noiseequality}) we
anticipate the following inequality to hold:
\begin{equation}
\label{noiseinequality} \overline{\chi^2}(|d/2,d/2\rangle,d,D) <
\overline{\chi^2}(|d\rangle_\phi,d,D),
\end{equation}
which completes the proof.

We have investigated validity of this inequality by comparing the
noise in numerically drawn histograms for the state
$|d/2,d/2\rangle$ and $|d\rangle_\phi$ but the observed difference
did not exceed the level of statistical error. This means that the
difference between $\overline{\chi^2}(|d/2,d/2\rangle,d,D)$ and $
\overline{\chi^2}(|d\rangle_\phi,d,D)$ is very small, which
reflects the fact that the probability of drawing the highly
asymmetric state in (\ref{approxdensity}) is negligible. The
non-intuitive inequality (\ref{noiseinequality}), although very
weak, must be an interesting signature of non-trivial spatial
correlations present within the mathematical structure of the Fock
states.

The inequality (\ref{noiseinequality}) can be seen also from the
structure of the analytic expression (\ref{symmetricdensity}). Let
us notice that when the number $d$ of the drawed particles
approaches its maximum value $2n$ then the width of the
distribution $\binom{2n-d}{n-d+j}$ equal to
$\frac{\sqrt{2n-d}}{2}$ rapidly shrinks. It follows that the more
particles we measure, the more symmetric states (which more likely
contribute to the interference) are being chosen for the drawing
of the positions.

Our last conclusion is that in the limit of $d\gg 1$, the
probability distribution $2^{-d}\binom{d}{j}$ from the
Eq.~(\ref{approxdensity}) becomes relatively narrow as
$\frac{\sqrt{d}}{2}\ll d$ and only the states $|j,d-j\rangle$ that
are almost symmetric will be chosen for the drawing of the
particle positions. Therefore in the large particle number limit
all quantities that weakly depend on the asymmetry of the state
will reproduce the results obtained for the phase states
$|d\rangle_\phi$.

We have proven the existence of the interference effects by
studying the structure of the high-order correlation functions for
the Fock states. It is also clear that these effects will
disappear after averaging out over many repetitions of the
measurement. This result is, however, an immediate consequence of
the Bogoliubov method which assumes {\em ad hoc} that one can
replace the field operator of a single condensate by a classical
wave with small quantum corrections: $\hat{\Psi}\approx
\sqrt{N}e^{i\phi} + \delta\hat{\Psi}$ and neglecting the latter.
Our analysis allows one to attribute the arbitrarily chosen phase
$\phi$ in the Bogoliubov method with the parameter $\phi$ from the
equation (\ref{approxdensity}) spontaneously induced in a single
measurement. In this interpretation breaking the phase-space
symmetry of the Fock states by using the Bogoliubov method
corresponds to replacing the strict expressions given by
Eq.~(\ref{symmetricdensity}) with their approximations
(\ref{approxdensity}).

We are mostly grateful to Krzysztof Sacha for letting us use his
numerical procedure for drawing particle positions. We would also
like to thank Jan Mostowski and Czes"{l}aw Radzewicz for valuable
comments on the manuscript and Konrad Banaszek and Wojtek
Wasilewski for useful discussions.

\end{document}